\documentclass[conference]{IEEEtran}
\IEEEoverridecommandlockouts
\usepackage{graphicx}
\usepackage{amssymb}
\usepackage{amsmath}
\usepackage{mathtools}
\usepackage{blkarray, bigstrut} 
\usepackage{cite}
\usepackage{amsthm}
\usepackage{epstopdf}
\usepackage{color}
\usepackage{caption}
\usepackage{subcaption}
\usepackage{soul}
\usepackage{balance}
\usepackage{multirow}
\usepackage{multicol}
\usepackage{booktabs}
\usepackage{amsfonts}
\usepackage{epsfig}
\usepackage{url}
\usepackage{algorithm}
\usepackage{algorithmic}
\usepackage{mathrsfs}
\usepackage{amsmath}
\usepackage{physics}
\usepackage{hyperref}

\usepackage{mathtools}

\def\mclimits_#1{\limits_{\mathclap{#1}}}

\usepackage[dvipsnames]{xcolor}
\definecolor{TungComment}{RGB}{0, 0, 255}
\definecolor{TungAsk}{RGB}{219, 48, 122}
\definecolor{TungFix}{gray}{0.6}
\definecolor{DuongFix}{RGB}{120, 255, 0}

\usepackage[english]{babel}
\usepackage[utf8]{inputenc}
\usepackage{textcomp}

\newtheorem{theorem}{Theorem}

\newtheorem{remark}{Remark}

\newtheorem{assumption}{Assumption}

\makeatletter
\def\ScaleIfNeeded{%
\ifdim\Gin@nat@width>\linewidth \linewidth \else \Gin@nat@width \fi
} \makeatother

\def\BibTeX{{\rm B\kern-.05em{\sc i\kern-.025em b}\kern-.08em
    T\kern-.1667em\lower.7ex\hbox{E}\kern-.125emX}}
\begin{document}
\title{Resource Allocation for Compression-aided Federated Learning with High Distortion Rate 
}


\author{\IEEEauthorblockN{Xuan-Tung~Nguyen${}^{\star}{}^{\diamondsuit}$, Minh-Duong Nguyen${}^{\blacklozenge}{}^{\diamondsuit}$, Quoc-Viet~Pham${}^{\blacklozenge}$, Quang-Vinh~Do${}^{\blacklozenge}$, Won-Joo~Hwang${}^{\blacklozenge}$}
    \IEEEauthorblockA{
    ${}^{\star}$~Sejong University, Republic of Korea \\
    ${}^{\blacklozenge}$~Pusan National University, Republic of Korea \\
    ${}^{\diamondsuit}$~Equal Contribution \\
    }
}

\maketitle

\begin{abstract}
Recently, a considerable amount of works have been made to tackle the communication burden in federated learning (FL) (e.g., model quantization, data sparsification, and model compression). However, the existing methods, that boost the communication efficiency in FL, result in a considerable trade-off between communication efficiency and global convergence rate. We formulate an optimization problem for compression-aided FL, which captures the relationship between the distortion rate, number of participating IoT devices, and convergence rate. Following that, the objective function is to minimize the total transmission time for FL convergence. Because the problem is non-convex, we propose to decompose it into sub-problems. Based on the property of a FL model, we first determine the number of IoT devices participating in the FL process. Then, the communication between IoT devices and the server is optimized by efficiently allocating wireless resources based on a coalition game. Our theoretical analysis shows that, by actively controlling the number of participating IoT devices, we can avoid the training divergence of compression-aided FL while maintaining the communication efficiency. 
\end{abstract}
\section{Introduction}
\label{sec:introduction}
Google introduced federated learning (FL) as a cost-effective approach for distributed machine learning without relying on a centralized data center \cite{2017-FL-First}. In general, FL enables devices to collaboratively learn a shared model without sending data from devices to a centralized server. As a result, considerable research has been conducted to implement the FL concept in practical situations, particularly in massive IoT networks.

Recently, many studies have been conducted regarding the implementation of FL for massive IoT networks \cite{2020-Survey-FL-MEC, 2021-FL-Survey-IoT}. The work in \cite{FL-2018-ConvergenceTime} proposed a joint distance-based user selection and resource block allocation algorithm to minimize the FL convergence time. In \cite{FL-2019-SchedulingPolicies}, the authors proposed different user scheduling policies for an FL system to enhance the FL convergence rate by reducing the number of communication rounds. In \cite{FL-2019-Optimization-AnalysisModel}, the authors addressed the trade-off between computation and communication efficiency considering the impact of various wireless communication factors (e.g., devices' power constraints and local data sizes). Therefore, they could optimize the total energy consumption of the FL training process under different settings for IoT devices. The research in \cite{2021-Comm-EnergyFL-Optimization} proposed a lower boundary for the number of FL training iterations. Consequently, they formulated a joint computation and transmission optimization problem to minimize total energy consumption for local computation and network communication. The authors in \cite{2021-FL-LearningRateOptimization-OTA} utilized a dynamic learning--rate method to reduce the aggregation error of FL in wireless networks. Nevertheless, these works focused on the vanilla FedAvg \cite{2017-FL-First}, in which no communication efficient method, such as model quantization \cite{2020-FL-TernaryFL} and model sparsification \cite{2020-FL-FedPAQ} methods, was integrated into the FL process. Therefore, the impact of lossy compression on FL is still under-explored. 

The main contribution of this work is the development of a novel framework that can support the FL process with high model compression by investigating the relationship between the FL convergence rate and model weight variance. \textit{To the best of our knowledge, this is the first work providing FL convergence--rate analysis under a distortion rate from model compression}. To this end, we propose a compression-aided FL model for IoT networks, in which a server located at the base station (BS) estimates the long-term communication load of the entire network, which is necessary for the FL convergence. We then formulate a joint resource-allocation and user-selection problem for FL aiming to minimize the total communication time. In this problem, we first consider the minimum number of participating IoT devices for the FL process, which must ensure the FL convergence.We then employ a coalition game to find an efficient bandwidth assignment for the proposed optimization problem. 

The rest of this article is organized as follows. In Section~\ref{sec:system-model}, we introduce the system model, fundamental knowledge about FL, and problem formulation. In Section~\ref{sec:convergence-rate}, we investigate the relationship between the FL convergence rate and other features of compression-aided FL. The proposed algorithm is presented in Section \ref{sec:proposed-algorithm}. Simulations are provided in Section~\ref{sec:performance-evaluation}. Finally, Section~\ref{sec:conclusion} concludes this paper.
\section{System Model and Problem Formulation}
\label{sec:system-model}
We consider a multi-IoT device single-server wireless network consisting of a set $\mathcal{K}=\{1,2,\dots,K\}$ of $K$ IoT devices and one base station (BS).For FL, the BS selects $K^\textrm{SE}$ IoT devices to participate in each FL communication round. These IoT devices and the BS jointly execute an FL algorithm for data analysis and model inference. The IoT devices collect from environment a set $\mathcal{D}=\{1,2,\dots,D\}$ of $D$ data samples and process them via deep neural networks. We assume that the data in our research are independent and identically distributed (IID). With this assumption, we can have $D = \sum^K_{k=1}{D_k}$, where $D_k$ is the number of data samples collected by the $k$-th IoT device each round.

In this paper, we use the Orthogonal frequency-division multiple access (OFDMA)  as the multiple-access scheme in the uplink. We assume that there are total $S$ sub-channels, the set of which is denoted by $\mathcal{S}=\{1,2,\dots,S\}$. We denote as $s$ the $s^\text{th}$ sub-channel. Each IoT device can only occupy one sub-channel. 
\subsection{Federated Learning}
\label{sec:system-model-sub:federated_learning}
In this work, we employ a cooperative training process between IoT devices and the server. Every participating IoT devices share the loss functions from their training operations to find the aggregated loss at the server. The global loss function at communication round $t$ is given by
\begin{equation}
\label{eq:FL_loss_function}
\begin{split}
\mathcal{F}(\boldsymbol{w}_g^t) = \frac{1}{K}{\sum_{k=1}^{K}} r_k f_k(\boldsymbol{w}_k^t),
\end{split} 
\end{equation}
 Here, $r_k \in\{0,1\}$ denotes the participation decision variable, where $r_{t,k}=1$ indicates that IoT device $k$ participates in the FL process and contributes the local model parameters for the aggregation at the server; otherwise, $r_{t,k}=0$. In \eqref{eq:FL_loss_function}, $\boldsymbol{w}_g^t$ denotes the global model parameters, $f_k(\boldsymbol{w}_k^t)$ denotes the loss function on IoT device $k$ after the $t$-th training round, and $K^\textrm{SE}$ is the number of selected IoT devices. It is noteworthy that different loss function is employed in different IoT devices. In an alternative way, we can demonstrate \eqref{eq:FL_loss_function} as follows:
\begin{equation}
\label{eq:FL_FedAvg}
\begin{split}
\boldsymbol{w}_g^t=\sum^{K}_{k=1}{\boldsymbol{w}_{k}^t \left(\frac{r_{k}D_k}{\sum_{j=1}^{K}{r_{j}D_j}}\right)}.
\end{split} 
\end{equation}
In each round, the server selects a fixed $K^\textrm{SE}$ IoT devices to join the FL training process. Hence, we have $\sum^{K}_{k=1}{r_{t,k}}=K^\textrm{SE}$.
\subsection{Lossy Federated Learning over Model Distortion}
\label{sec:convergence-rate}
In the concept of compression-aided FL, the trained model in each IoT device will be compressed before being transmitted to enhance communication efficiency. However, using lossy compression (i.e., the data is compressed at an extremely high rate) may cause considerable information loss (i.e., compressing and decompressing cause the data to progressively lose quality). To be more specific, information loss can be understood as the reduction in mutual information $I(\boldsymbol{w}_k;\hat{\boldsymbol{w}}_k)$ between the original and decompressed model on devices $k$ \cite{MF-2006-InformationTheory}. This phenomenon leads to the distortion in data reconstruction, which can be represented by the mean distance between every data point of the original and decompressed models, as $d(\boldsymbol{w}_k;\hat{\boldsymbol{w}}_k) = \mathbb{E}_{i\in \abs{\boldsymbol{w}_k}}d(w_{i,k};\hat{w}_{i,k})$. Here, the distance can be measured using various approaches (e.g., a mean square error approach) \cite{MF-2006-InformationTheory}. Thereby, the error between the original and reconstructed models leads to the lossy-FL flop compared to the conventional FL methods (e.g., FedAvg) due to model modification, which cause training divergence in the FL process. 

Nevertheless, the work in \cite{FL-2021-HCFL} proved that increasing the number of participating IoT devices can greatly plunge the training error, thus, achieve a faster convergence rate. Observing that, we aim to formulate the problem that can find an equilibrium at which both the communication efficiency and convergence rate are satisfied. To achieve this equilibrium, we propose a function that estimates the relationship among the number of participating IoT devices, model distortion (owing to the lossy compression scheme), and the FL convergence rate.
\subsubsection{Distortion Effect in Federated Learning}
The compression-integrated FL method always generates distortion between the original model parameters and the model reconstructed from the compressed data \cite{MF-2006-InformationTheory}. Specifically, the reconstructed model at the IoT devices when applying a compression scheme can be represented as follows:
\begin{align}
    \boldsymbol{w}_k = \hat{\boldsymbol{w}}_k + \sigma
\end{align}
As proven in \cite{FL-2021-HCFL}, the reconstructed model bias follows a Gaussian distribution with a standard deviation of $\sigma$. The relationship between the deviation and other FL settings, such as the compressor's reconstruction loss $\mathcal{L}(w)$ and the number of participating IoT devices $K^\textrm{SE})^2$, are given by
\begin{equation}
\label{eq:FL_model_bias}
\begin{split}
\sigma^2 \leq \frac{2}{( K^\textrm{SE})^2}\mathcal{L}(w).
\end{split} 
\end{equation}
As observed from \eqref{eq:FL_model_bias}, the standard deviation $\sigma$ is inversely proportional to $(K^\textrm{SE})^2$. Intuitively, as the number of participating IoT devices increases, the reconstructed model parameters is dramatically getting closer to their original values. 
\subsubsection{Assumptions and Notations}
\label{sec:problem-formulation:assumptions-notations}
To evaluate the FL performance, we make the following assumptions for the local loss functions at IoT devices $f_1, \dots, f_K$.
\begin{assumption}
    \textit{$f_1, f_2, \dots, f_K$ are $L$-smooth $\forall p,q$, $f_k(p)\leq f_k(q)+(p-q)^T \nabla f_k(q) + \frac{L}{2}\norm{p-q}^2_2$}.
\end{assumption} 
\begin{assumption}
    $f_1, f_2, \dots, f_K$ are $\mu$-strongly convex $\forall p,q$, $f_k(p)\geq f_k(q)+(p-q)^T \nabla f_k(q) + \frac{\mu}{2}\norm{p-q}^2_2$.
\end{assumption}
\begin{assumption}
    Assuming that the $k$-th device samples $\xi^k_t$ dataset from their domain ($\abs{\xi^k_t} = D_k$) at iteration $t$, we have the upper-boundary for the stochastic gradient's variance per device as $\mathbb{E}\norm{\nabla f_k(\boldsymbol{w}^t_k, \xi^k_t) - \nabla f_k(\boldsymbol{w}^t_kl)}^2 \leq \frac{1}{K_\textrm{SE}}\sum^{K_\textrm{SE}}_{k=1}\delta^2_k$.   
\end{assumption}
\begin{assumption}
    For all $ k \in \{1,2,\dots, K\}$ as devices' indices, and at any iteration $t= \{0,2, \dots, R\}$, we have the following boundary for the stochastic gradient: $\mathbb{E}\norm{\nabla f_k(\boldsymbol{w}^k_t,\xi^k_t)}^2 \leq G^2$.   
\end{assumption} 
\subsubsection{Convergence Rate Analysis}
\label{sec:problem-formulation:convergence-result}
\begin{theorem}
Let Assumption 1 to 4 hold and $G, \delta_k, \mu, L$ are defined inward, we define the following notations:
\begin{align}
    \chi =~&\sum^K_{k=1}{\Big[f_k(\boldsymbol{w}^k_t)-f_k(\boldsymbol{w}^*)\Big]},  \label{eq:Theorem1-Notation4}\\
    U_1 =~&\frac{\mathcal{L}(w)}{(K^\textrm{SE})^2(E+1)}, \label{eq:Theorem1-Notation1} \\
    U_2 =~&\frac{1}{\mu Ea}\Big(E6L\chi + 8E(E-1)^2 G^2 + \frac{1}{K^\textrm{SE}}\sum^{K_\textrm{SE}}_{k=1}\delta^2_k\Big) \notag\\
    &+ \frac{\mathcal{L}(w)}{(K^\textrm{SE})^2(E+1)}, \label{eq:Theorem1-Notation2} \\  
    U_3 =~&\frac{a\mathcal{L}(w)}{(K^\textrm{SE})^2(E+1)} - \frac{4a}{\mu^2}G^2+ \notag\\
      &\frac{2}{\mu E}\Big(E6L\chi + 8E(E-1)^2 G^2 + \frac{1}{K^\textrm{SE}}\sum^{K_\textrm{SE}}_{k=1}\delta^2_k\Big) \label{eq:Theorem1-Notation3},
\end{align}
\textit{where $E$ is the number of epochs required for local training. The subscript $\chi$ represents the expected distance between local loss on each IoT device and the global loss of the FL system.
By choosing learning rate $\eta_t = \frac{E+1}{E\mu(a+t)}$,  $a = \max \{\frac{8L}{\mu},E\}$, we have the following theorem: }
\begin{equation}
\label{eq:Theorem1-CommunicationTime}
\begin{split}
        R &\geq R^{\min}(K^\textrm{SE}) \\ &=\left\lceil\frac{(\frac{2 \epsilon}{L}-U_2)+\sqrt{(U_2-\frac{2 \epsilon}{L})^2+4U_1U_3}}{2U_1E}\right\rceil,
\end{split} 
\end{equation}
where $\lceil \cdot \rceil$ denotes the ceiling operator. $R^{\min}$ is the lower boundary for the total communication time of the proposed algorithm.
\end{theorem}

\textit{Proof.} Due to the space limitation, the proof is omitted.
\subsection{Communication Model}
\label{sub-2:communication_model}
In this section, we discuss the communication delay model in a FL-integrated wireless network. 
We apply OFDMA for the uplink data transmission. In this network, each IoT device can only occupy one sub-channel. Thus, the data rate of the $k$-th IoT device at communication round $t$ via the $s$-th sub-channel is given by
\begin{equation} 
\label{eq:CM_rate_uplink}
\begin{split}
c_{k}(\boldsymbol{\psi}_{t,k}) =\sum^{S}_{s=1}\psi_{t,k,s}B\log_2{\left(1+\frac{P_{k}|h_{t,k,s}|^2}{BN_0}\right)},
\end{split} 
\end{equation}
where $r_{t,k}=1$ indicates that the $k$-th IoT device is selected to upload the data to the server in round $t$; $r_{t,k}=0$ otherwise. Meanwhile, $\psi_{t,k,s}=1$ indicates that the $k$-th IoT device occupies the $s$-th sub-channel in round $t$; $\psi_{t,k,s}=0$ otherwise. Additionally, $h_{t,k,s}$ denotes channel power gain, $P_k$ denotes the transmit power, $B$ is the channel bandwidth, and $N_0$ is the noise power spectral density. Given the achievable rate of all IoT devices, $c_{k}(\psi_{t,k,s})$, we have the total transmission time in the $t$-th communication round of the model update from IoT device $k$ to server as follows: 
\begin{equation}\label{eq:CM_delay_saving}
\begin{split}
T_{t}^\textrm{COMP}(\boldsymbol{r}_t,\boldsymbol{\Theta}_t, K^\textrm{SE}) =  Z_t^{\textrm{COMP}} \sum^{ K}_{k=1} r_{t,k}\frac{1}{c_k},
\end{split} 
\end{equation}
where $\boldsymbol{\Theta}_t=[\boldsymbol{\psi}_{t,1},...,\boldsymbol{\psi}_{t,K}]$. The size of the compressed model is denoted by $Z^{\textrm{COMP}}$. In the general FL setting \cite{2017-FL-First}, IoT devices are assumed to use the same model for convenient FL implementation. Thus, the same $Z^{\textrm{COMP}}$ value is applied on every device. By calculating the total transmission time in the $t$-th communication round, we formulate the long-term transmission time minimization problem, as presented in Section~\ref{subsec:problem-formulation}.
%
\subsection{Problem Formulation}
\label{subsec:problem-formulation}
We concentrate on the total communication time for the compression-integrated FL process. Our objective is to minimize the model transmission time to achieve the desired global accuracy. Specifically, the transmission time is calculated by summing up the transmission time from $R$ communication rounds. In which, the transmission time at the $t$-th round is calculated independently as in \eqref{eq:CM_delay_saving}. Normally, in various works (e.g.,\cite{2022-FL-OptimizeWirelessIoTNetworks, 2021-FL-JointCommFramework}), the minimum transmission time is estimated based on the loss function of the FL. However, the lack of understanding of FL's behavior makes the problem difficult to obtain the solution, leading the problem to be prone to overfitting. To make the problem feasible, we apply the Theorem~$1$ (i.e., we find the optimal lower bound on $R$ so that the total transmission time is minimized). Intuitively, by reducing the lower bound, we can reduce the expected value of the probability distribution of $R$ value. As a consequence, $R$ has a high chance to decrease as the lower bound $R_\textrm{min}$ is reduced. For convenience, we define $\boldsymbol{R}=[\boldsymbol{r}_1,...,\boldsymbol{r}_R]$, and $\boldsymbol{\Phi}=\{\boldsymbol{\Theta}_1,...,\boldsymbol{\Theta}_R\}$. The optimization problem is mathematically formulated as:
\begin{subequations}
	\label{subeqn-opt-pro-general:opt-pro-main}
	\begin{alignat} {3}
		& \min_{\boldsymbol{R}, \boldsymbol{\Phi}, K^{\textrm{SE}}} 
		&	 &  \sum^{R^{\min}(K^\textrm{SE})}_{t=1} T_t^\textrm{COMP}(\boldsymbol{R},\boldsymbol{\Theta}_t, K^\textrm{SE}) 
		\label{subeqn-opt-pro-general:opt-pro} \\
		& \text{s.t.}
		&	& 0< K^\textrm{SE} \leq K,
		\label{subeqn-opt-pro-general:number-IoT devices-per-round}\\ 
		&   &   & \sum^{K}_{k=1}{r_{t,k}} = K^\textrm{SE}, \forall t\in  [1,...,R], r_{t,k}=\left\{0,1\right\},
		\label{subeqn-opt-pro-general:total-participating-IoT device}\\ 
        &	&	&  \sum_k^{K} \psi_{t,k,s} =  1, \forall t, \forall s, \psi_{k,t,s} = \left\{0,1\right\},
		\label{subeqn-opt-pro-general:number-IoT devices-per-subchannel}\\
		&	&	&  \sum^{S}_{s=1}\psi_{t,k,s} \geq 1, \forall k \in \mathcal{K}_t^{\textrm{SE}}, \forall t\in [1,..,R^{\min}(K^\textrm{SE})] 
		\label{subeqn-opt-pro-general:total-usage-bandwidth},\\
		&	&	&  \sum^{K}_{k=1}\sum^{S}_{s=1}\psi_{t,k,s} = S, \forall t\in [1,..,R^{\min}(K^\textrm{SE})] 
		\label{subeqn-opt-pro-general:total-bandwidth},
    \end{alignat}
\end{subequations}
where \eqref{subeqn-opt-pro-general:number-IoT devices-per-round} is the feasible condition for the number of participating IoT devices. Meanwhile, \eqref{subeqn-opt-pro-general:total-participating-IoT device} indicates that the number of participating IoT devices in communication round $t$ is equal to $K^\textrm{SE}$. \eqref{subeqn-opt-pro-general:number-IoT devices-per-subchannel} indicates that each sub-channel is occupied by at most one IoT device in each communication round. \eqref{subeqn-opt-pro-general:total-usage-bandwidth} indicates that one IoT devices can use more than one sub-channel. \eqref{subeqn-opt-pro-general:total-bandwidth} is used to ensure that all sub-channels will be utilized.
\section{Proposed Algorithm}
\label{sec:proposed-algorithm}
From \eqref{subeqn-opt-pro-general:opt-pro-main}, we can observe that the objective function depends on integer variables (i.e., $\boldsymbol{r}_{t},\boldsymbol{\psi}_{t,k},K^\textrm{SE}$), and each variable is limited by the constraints. With limited searching space, the optimal solution can be obtained by exhaustive search algorithm. However, the exhaustive search is impossible since the complexity is $\mathcal{O}\left(K \textrm{C}^{K^{\textrm{SE}}}_K \textrm{P}^{K^{\textrm{SE}}}_S(S-K^{\textrm{SE}})!\right)$
\footnote{$\textrm{C}^{K^{\textrm{SE}}}_K=\frac{K!}{K^{\textrm{SE}}!(K-
K^{\textrm{SE}})!}$, $\textrm{P}^{K^{\textrm{SE}}}_S=\frac{S!}{(S-
K^{\textrm{SE}})!}$.}
, which increases significantly when the network size increases. Besides, we notice that the objective function is the sum of $R^{\min}(K^\textrm{SE})$ elements. Meanwhile $R^{\min}(K^\textrm{SE})$, the number of communication rounds, plays a vital role in this problem. Specifically, the number of communication rounds not only impacts on the total communication time, but also the computational requirements. Furthermore, $R^{\min}(K^\textrm{SE})$ only depends on the number of participating IoT devices per round, $K^\textrm{SE}$. Inspiring from the above justifications, we suggest to solve the problem in \eqref{subeqn-opt-pro-general:opt-pro-main} by firstly optimizing the total number of communication rounds, $R_{\textrm{min}}$, with respect to the number of selected IoT devices per round, $K^\textrm{SE}$. We then optimize the total time consumption for each round, $T_{t}^\textrm{COMP}(\boldsymbol{r}_t,\boldsymbol{\Theta}_t)$, with respect to the selection of selected IoT devices, $\boldsymbol{r}_t$, and the sub-channel assignment, $\boldsymbol{\Theta}_t$.
\begin{algorithm}[h]
	\caption{Optimal number of participating IoT device search} \label{Algorithm1}
	\begin{algorithmic}[1]
		\STATE \textbf{Input}: {Set $k=1$, and calculate the number of communication round $R_1^{\textrm{min}}$ as in \eqref{eq:Theorem1-CommunicationTime}.}
		\STATE {Set $K^{\textrm{SE,opt}}=1$}
		\FOR {$k \in [2:K]$}
		\STATE {Calculate $R_k^{\textrm{min}}$ as in \eqref{eq:Theorem1-CommunicationTime}}.
		\IF {$R_k^{\textrm{min}}<R_{k-1}^{\textrm{min}}$}
		\STATE {Set $K^{\textrm{SE,opt}}=k$}
        \ENDIF 
		\ENDFOR
		\STATE {\textbf{Output}: The optimal number of selected IoT devices per communication round $K^{\textrm{SE,opt}}$}.
	\end{algorithmic}
\end{algorithm} 
\subsection{Communication rounds minimization problem}
In this subsection, we aim to optimize the number of communication rounds under the constraint of the number of selected IoT devices per round. The problem is given by 
\begin{subequations}
	\label{subeqn-opt-pro-general:opt-R-main-sub1}
	\begin{alignat} {3}
		& \min_{K^\textrm{SE}} ~~~
		&	 &  R^{\min}(K^\textrm{SE})
		\label{subeqn-opt-pro-general:opt-R}\\ 
		& \text{s.t.} 
		&	& 0< K^\textrm{SE} \leq K.
	\end{alignat}
\end{subequations}
We observe that this problem is non-convex even after relaxing the discrete variable to continuous one. However, since the complexity of the problem increases linearly with the increase in the number of IoT devices, the optimal number of IoT devices per round can be found by searching over all possible candidates as described in Algorithm~\ref{Algorithm1}.
\subsection{Communication time consumption minimization}
Given the number of selected IoT devices per round, our mission is to choose the IoT devices and assign sub-channels to them for model uploading. The problem of minimizing the total transmission time at the $t$-th communication round can be written as
\begin{subequations}
	\label{subeqn-opt-pro-general:opt-pro-main-sub2}
	\begin{alignat} {3}
		& \min_{\boldsymbol{r}_t,\boldsymbol{\Theta}_t} ~~~
		&	 &  T_{t}^\textrm{COMP}(\boldsymbol{r}_t,\boldsymbol{\Theta}_t) 
		\label{subeqn-opt-pro-general:opt-pro-sub2}       \\
		& \text{s.t.}
		&	& \eqref{subeqn-opt-pro-general:total-participating-IoT device},
		\eqref{subeqn-opt-pro-general:number-IoT devices-per-subchannel}, \eqref{subeqn-opt-pro-general:total-usage-bandwidth}, \eqref{subeqn-opt-pro-general:total-bandwidth}.
	\end{alignat}
\end{subequations}
Occasionally, the device selection variable $\boldsymbol{r}_{t}$ is utilized to optimize the energy consumption in FL. However, the locations of IoT devices are geographically fixed, which can leads to the static distribution of devices' channel gain. Therefore, the optimum value of $T_{t}^\textrm{COMP}(\boldsymbol{r}_t,\boldsymbol{\Theta}_t)$ is lack of randomness, and may lead to the bias in choosing participating IoT devices for the FL's communication round. Thus, the FL loses the generalization over all of the dataset (i.e., the lack of data sampling over the whole network). Thus, we apply a random selection scheme for the optimization problem in \eqref{subeqn-opt-pro-general:opt-pro-main-sub2} to obtain $\boldsymbol{r}_{t}$. The problem in \ref{subeqn-opt-pro-general:opt-pro-main-sub2} is then optimized with respect to $\boldsymbol{\Theta}_t$ as follows:
\begin{subequations}
	\label{subeqn-opt-pro-general:opt-pro-main-sub3}
	\begin{alignat} {3}
		&\min_{\boldsymbol{\Theta}_t} ~~~
		&	 &  T_{t}^\textrm{COMP}(\boldsymbol{\Theta}_t) =Z_t^{\textrm{COMP}} \sum^{ K}_{k=1} r_{t,k}\frac{1}{\sum^{S}_{s=1}\psi_{t,k,s}c_{t,k,s}} 
		\label{subeqn-opt-pro-general:opt-pro-sub3}       \\
		& \text{s.t.}
		&   & \eqref{subeqn-opt-pro-general:number-IoT devices-per-subchannel}, \eqref{subeqn-opt-pro-general:total-usage-bandwidth}, \eqref{subeqn-opt-pro-general:total-bandwidth}.
	\end{alignat}
\end{subequations}
where $c_{t,k,s}=B\log_2{\left(1+\frac{P_{k}|h_{t,k,s}|^2}{BN_0}\right)}$ is the achievable rate of the $k$-th IoT device using the $s$-th sub-channel. Recall that the uplink achievable rate of an IoT device depends only on its sub-channel selection. However, the problem in \eqref{subeqn-opt-pro-general:opt-pro-main-sub3} is still difficult to solve because of the large number of available sub-channel and number of IoT devices. In this case, a coalition game method can be used to provide an efficient sub-channel assignment  with close optimal solution compared to the exhaustive search algorithm. To be specific, the problem in \eqref{subeqn-opt-pro-general:opt-pro-main-sub3} can be seen as a trading game among IoT devices in the system. In this game, IoT devices share a set of sub-channels, and each device selects sub-channel considering the sub-channel selections of other devices to obtain a common target, which is the minimum communication time, $T_{t}^\textrm{COMP}(\boldsymbol{\Theta}_t)$. We denote the set of sub-channels assigned to the $k$-th IoT device as $\mathcal{S}_k$, and we have $\mathcal{S}=\{\mathcal{S}_1,...,\mathcal{S}_K\}$. Following that, the sub-channels will be exchanged between subsets, $\mathcal{S}_k, \forall k\in\mathcal{K}$, to achieve the sub-optimal sub-channel assignment structure for the system. However, a sub-channel can be changed from one IoT device to another only when the channel exchange  provide lower total time consumption. 

Note that, each IoT device needs at least one sub-channel to communicate with the server. Therefore, when an IoT device has just one sub-channel, the controller  will exchange the sub-channel of this IoT device with other IoT device instead of moving the sub-channel.
The sub-channel assignment based on the coalition game method is described in Algorithm \ref{CoaltionGameAlgorithm}. 
Algorithm \ref{CoaltionGameAlgorithm} starts by initializing the sub-channel assignment structure. The controller checks over all sub-channel in the system in turn. Specifically, the number of sub-channels assigned to the same IoT devices is firstly verified (as in line $7$). This step is used to decide whether the sub-channel should be used by another device (line $10$) or should be exchanged with other sub-channel from another IoT device (line $8$). Next, if the switch of the location of the considered sub-channel can offer better performance to the system (line $10$), we then form a new sub-channel structure.
\begin{algorithm}[!h]
	\caption{Coalition game based method} \label{CoaltionGameAlgorithm}
	\begin{algorithmic}[1]
		\STATE \textbf{Input}: Initialize  sub-channel to all IoT devices satisfying \eqref{subeqn-opt-pro-general:number-IoT devices-per-subchannel}, \eqref{subeqn-opt-pro-general:total-usage-bandwidth}, and \eqref{subeqn-opt-pro-general:total-bandwidth}.
		\STATE {Denote the current sub-channel assignment structure as $\mathcal{S}^{\textrm{cur}}=\{\mathcal{S}^{\textrm{cur}}_k,\forall k\in \mathcal{K}\}$}
		\REPEAT
		\FOR {$s=1:S$}
		\FOR {$k=1:K$}
		\STATE {Assume that $s\in \mathcal{S}_j, j\in \mathcal{K}$}
		\IF {$|\mathcal{S}_k|==1$}
		\STATE {randomly choose a sub-channel, $s'\in \mathcal{S}_k$, and execute the switch to form a new structure as: 
		$\mathcal{S}^{\textrm{temp}}_j=\mathcal{S}^{\textrm{cur}}_j\setminus s \cup s'$ and $\mathcal{S}^{\textrm{temp}}_k=\mathcal{S}^{\textrm{cur}}_k\setminus s' \cup s$}
		\ELSE
		\STATE {$s$ leaves $\mathcal{S}_j$ to join $\mathcal{S}_k$, $\mathcal{S}^{\textrm{temp}}_k=\mathcal{S}^{\textrm{cur}}_k \cup s$} 
		\ENDIF 
		\IF {$T_{t}^\textrm{COMP}(\mathcal{S}^{\textrm{cur}})\geq T_{t}^\textrm{COMP}(\mathcal{S}^{\textrm{temp}})$} 
		\STATE {$\mathcal{S}^{\textrm{cur}}=\mathcal{S}^{\textrm{temp}}$}
		\ENDIF 
		\ENDFOR
		\ENDFOR
		\UNTIL {No IoT device want to exchange its time slot}
		\STATE \textbf{Output}: The sub-channel assignment to all IoT devices in the $t$-th communication round. 
	\end{algorithmic}
\end{algorithm} 
\begin{remark}[{{Convergence and complexity analysis}}]
    Since the number of sub-channels and IoT devices are constrained, the number of sub-channel candidates is also limited. Therefore, Algorithm \ref{CoaltionGameAlgorithm} will converge after certain iterations. Besides, assuming that Algorithm \ref{CoaltionGameAlgorithm} converses after $N^{\textrm{CG}}$ iterations. At each iteration, the system requires total $K^{\textrm{SE}}$ times of data rate calculation. Therefore, the complexity of Algorithm~\ref{CoaltionGameAlgorithm} is $\mathcal{O}(N^{\textrm{CG}}K^{\textrm{SE}})$. Similarly, the complexity of the exhaustive search is $\normalfont{\textrm{P}}^{K^{\textrm{SE}}}_S (S-K^{\textrm{SE}})!K^{\textrm{SE}}$.
\end{remark}

\section{Performance Evaluation} \label{sec:performance-evaluation}
In this section, we evaluate the performance of our proposed methods under various scenarios. Firstly, we analyze how the number of participating IoT devices affects to the model convergence rate. Then, we explain why the value of $R^\textrm{min}$ is chosen in problem~\ref{subeqn-opt-pro-general:opt-R-main-sub1}. Next, to demonstrate the effectiveness of our methods, we make comparisons with several benchmarks. We compare the proposed sub-channel assignment algorithm to the bandwidth fairness method, in which the controller  allocates almost the same amount of bandwidth to IoT devices without considering the channel conditions. The considered communication area is a circle with radius $r$ m, in which $K$ IoT devices are distributed randomly. Several important parameters of the system is provided in Table \ref{Table1}. \begin{table}[h]
		\caption{System parameters.}
		\centering
		\begin{tabular}{l|c}
			\toprule[1pt]\midrule[0.3pt] 
			Cell radius & 200 [m]\\ 
			\midrule 
			Transmit power of each IoT device & 23 [dBm]\\ 
			\midrule 
			Bandwidth & 180 [kHz]\\ 
			\midrule 
			Power spectral density of the thermal noise & -174 [dBm/Hz]\\ 
			\bottomrule[0.5pt] 
		\end{tabular}
		\label{Table1}
	\end{table}%

    \subsection{Evaluation on convergence rate theorem}
\begin{figure}[h]
    \centering
    \begin{subfigure}[b]{0.48\linewidth}
        \centering
        \includegraphics[width=\linewidth]{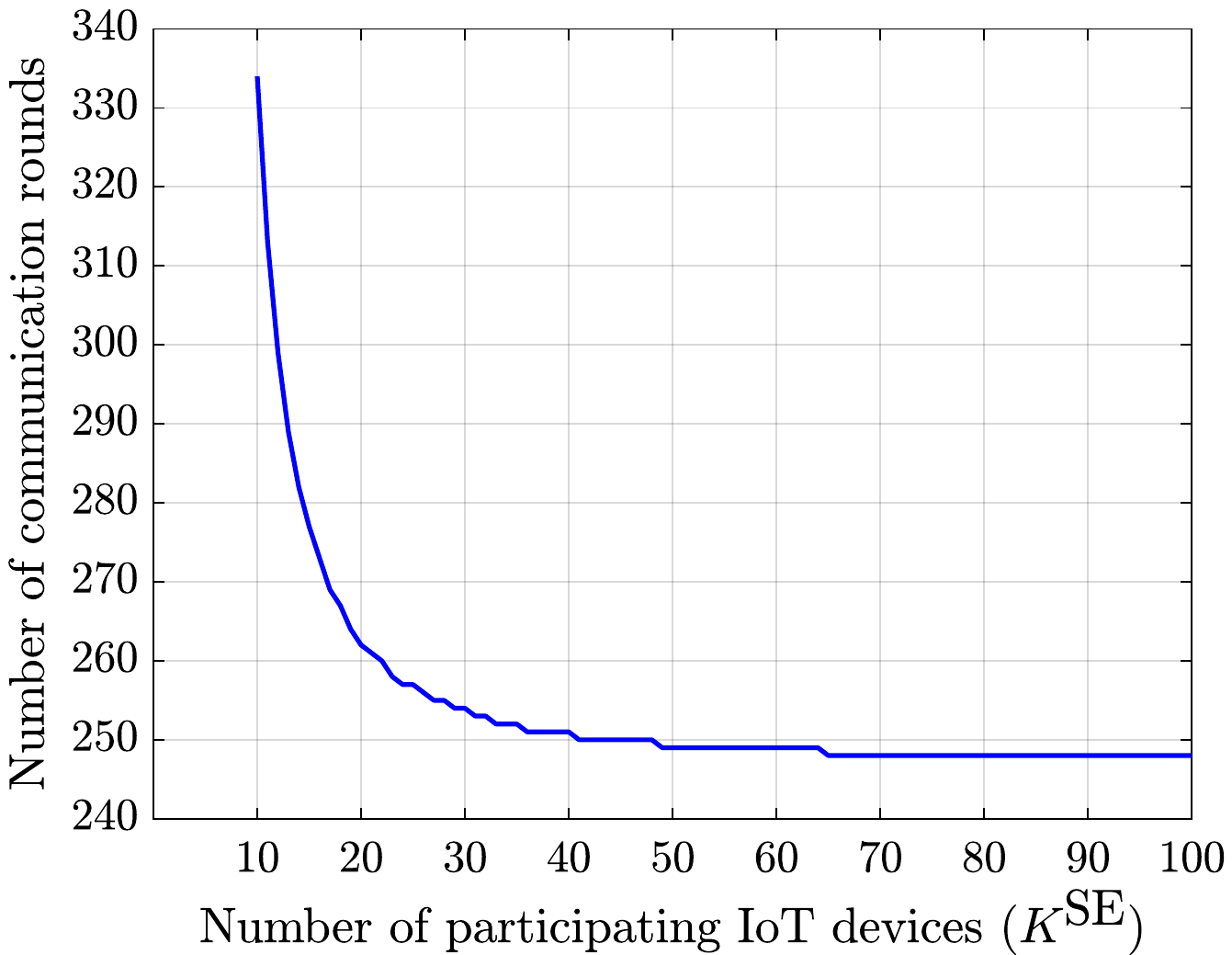}
        \caption{}
        \label{fig:impact-participateuser-convergencerate}
    \end{subfigure}
    \begin{subfigure}[b]{0.48\linewidth}
        \centering
        \includegraphics[width=\linewidth]{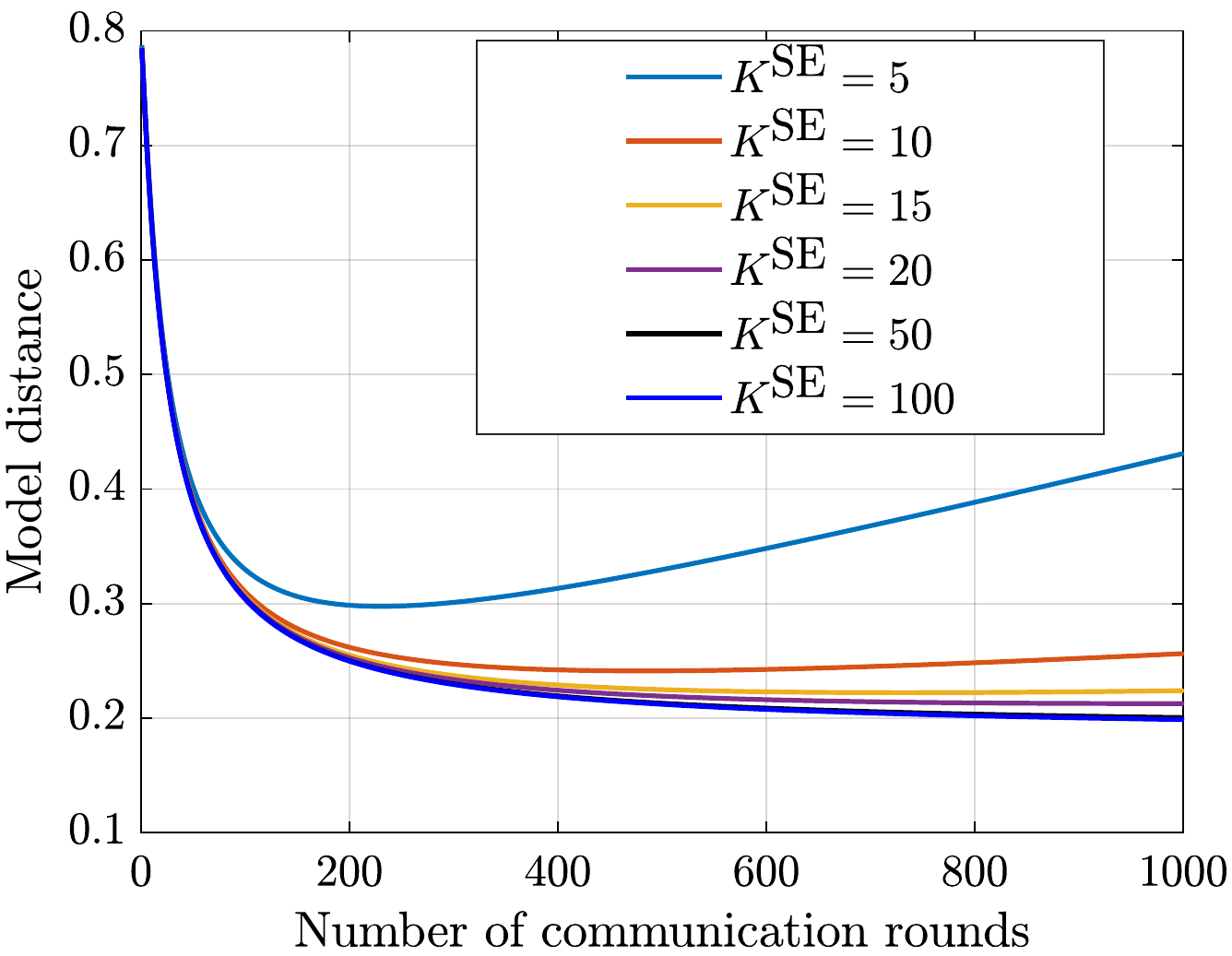}
        \caption{}
        \label{fig:impact-participateuser-commRound}
    \end{subfigure}
    \caption{Impact of different participating users on model convergence rate.}
    \end{figure}

    Figure~\ref{fig:impact-participateuser-convergencerate} illustrates the convergence rate of the FL algorithm when applying different numbers of participating devices. As can be seen from the figure, when the number of participating devices is low (i.e., $K^\textrm{SE} = 5$), the FL process tends to diverge after first $200$ rounds of global aggregation. The reason is that the FL process is affected by various distortion rates of the lossy compression at the devices. When $K^\textrm{SE}$ becomes large, the FL process converges efficiently to the optimal model. Nevertheless, when $K^{SE}$ is large enough, the convergence rate shows a tendency to stabilize (e.g., there is no significant difference in convergence rate between $K^\textrm{SE} = 50$ and $K^\textrm{SE} = 100$). Thus, by choosing a suitable value of $K^{SE}$, we can achieve the desired convergence rate while reducing the communication burden.
    To choose an appropriate $K^\textrm{SE}$, we observe the convergence of the FL process when changing $K^\textrm{SE}$ as, as illustrated in Figure~\eqref{fig:impact-participateuser-commRound}). Figure~\eqref{fig:impact-participateuser-commRound} reveals the number of communication rounds needed to achieve a certain accuracy for FL (i.e., when the distance between the current model to the optimal model is less than $0.2$). As we can see from the figure, when we use more than $40$ devices to participate each communication round, we can achieve the convergence in approximately 250 rounds. As a consequence, $K^\textrm{SE}$ can be chosen from $37$ to $40$ as a nearly optimum. With the chosen $K^\textrm{SE}$, we can ensure that the FL process can converge in the optimal time. Therefore, we only need to focus on resource allocation to achieve the optimal transmission time.
	\subsection{Evaluation on coalition game based method}
   
\begin{figure}[h]
    \centering
    \begin{subfigure}[b]{0.48\linewidth}
        \centering
        \includegraphics[width=\linewidth]{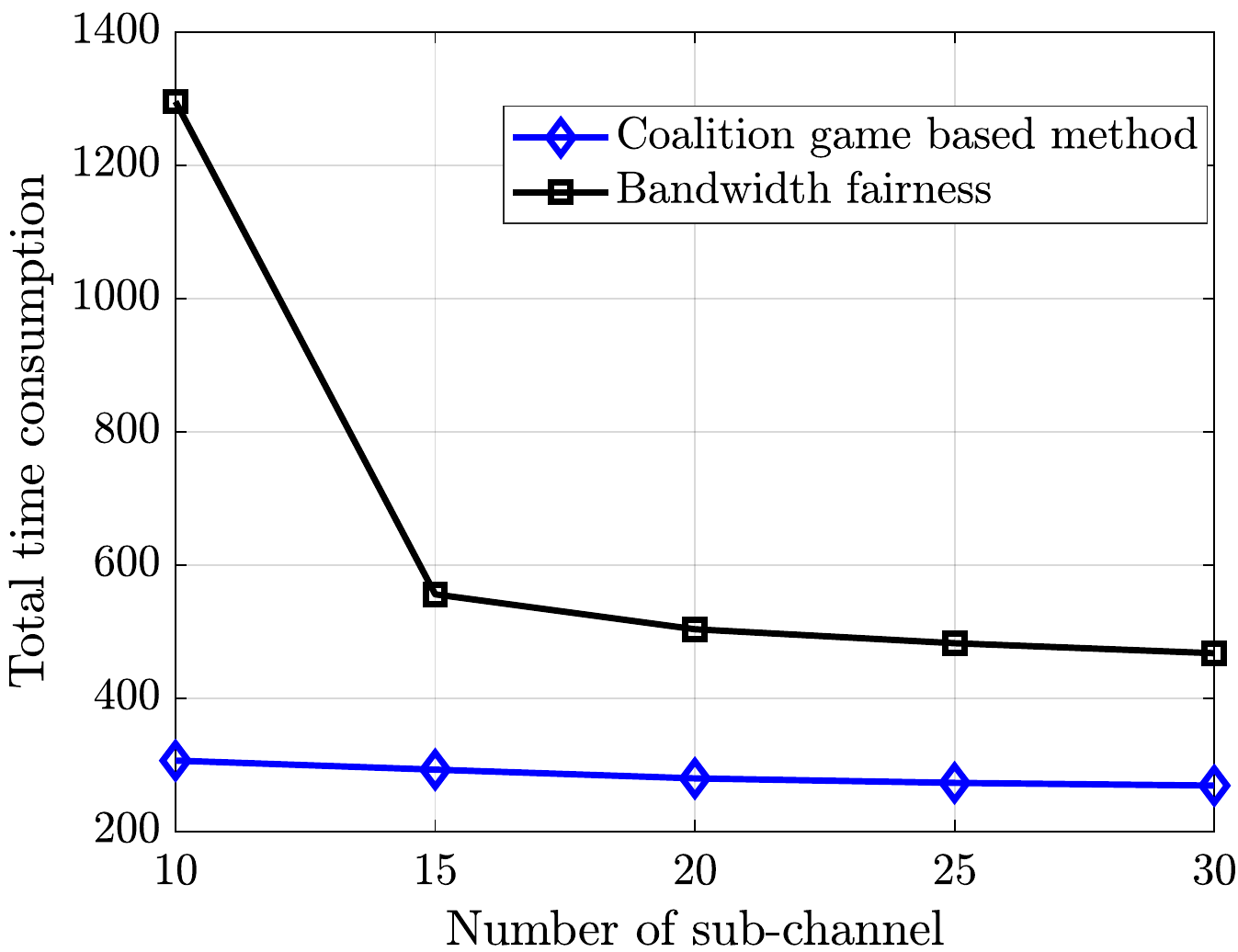}
        \caption{}
        \label{fig:increaseSubs}
    \end{subfigure}
    \begin{subfigure}[b]{0.48\linewidth}
        \centering
        \includegraphics[width=\linewidth]{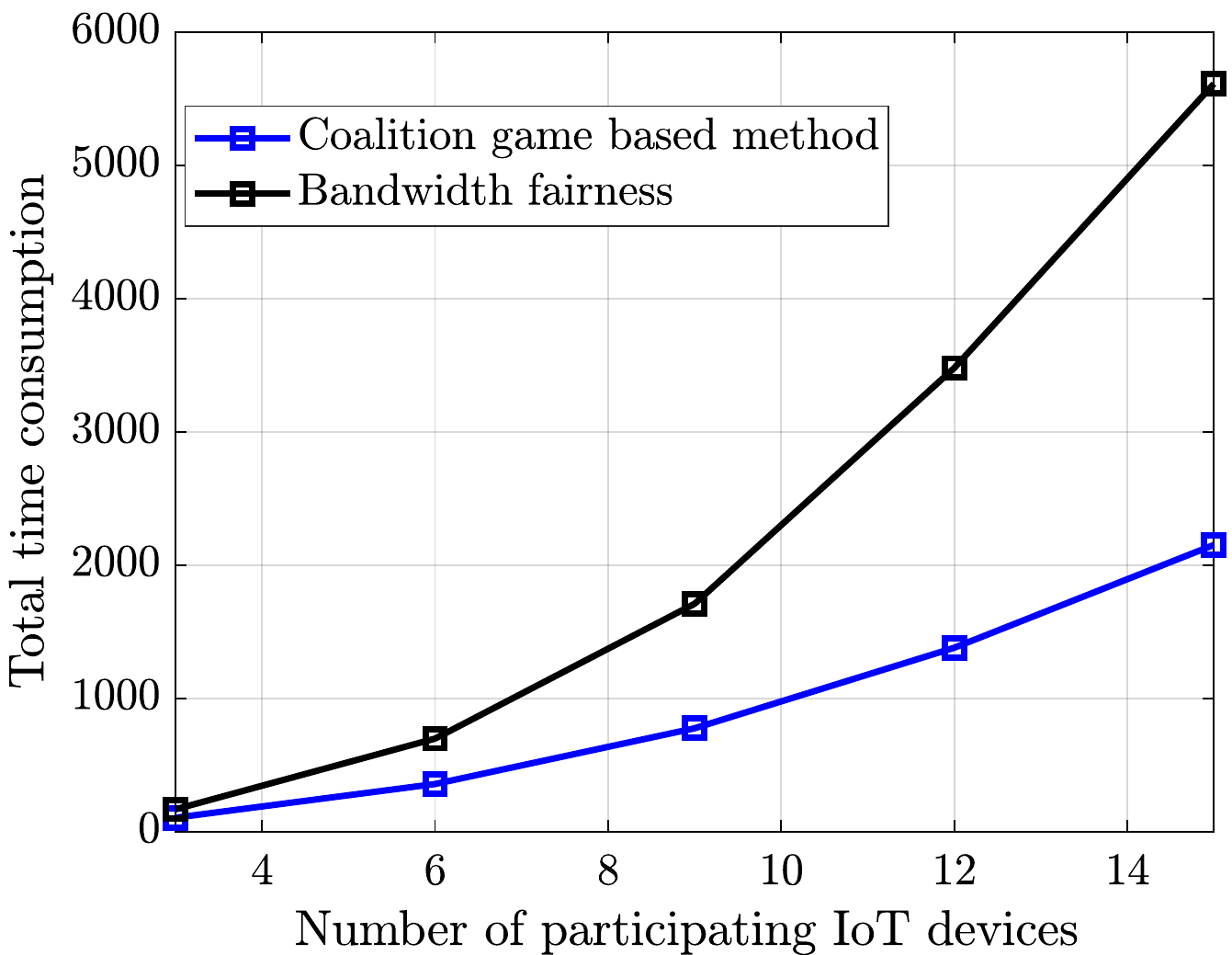}
        \caption{}
        \label{fig:increaseK}
    \end{subfigure}
    \caption{Total transmission time versus the number of sub-channels and number of participating IoT devices.}
    \label{fig:coalition}
    \end{figure}

    Figures~\ref{fig:coalition} describes the change of total transmission time when the network size increases. To be specific, after obtaining the number of participating IoT devices per communication round, the controller  will randomly select a set of IoT devices to upload their models to the server. The final results were obtained by averaging $10000$ simulations in . From Figure \ref{fig:increaseSubs}, there is a downward trend in the transmission time of both the proposed method and the bandwidth-fairness scheme. The reason is that increasing the number of sub-channels provides more bandwidth to IoT devices, thus, it increases the achievable rates and reduce transmission time. However, the proposed coalition game-based method can significantly reduce the transmission time, especially when the system has a small number of sub-channels. Specifically, the proposed method only needs $300$ seconds to transmit all data when the number of sub-channels is set to $10$. Meanwhile, the bandwidth fairness requires $1300$ seconds. Even when the number of sub-channels increases to $30$, the transmission time of the proposed method is around $250$ seconds, and that of the bandwidth-fairness scheme is almost double, $480$ seconds.
	The effectiveness of the coalition game-based method can be seen in Figure \ref{fig:increaseK}. The transmission time increases with the rise in the number of participating IoT devices. Following that, the transmission time of the proposed method increases linearly with the increase in the number of participating IoT devices ($K^\textrm{SE}$), from $50$ seconds at $K^\textrm{SE}=3$ to $2100$ seconds at $K^\textrm{SE}=15$. Meanwhile, the transmission time of the bandwidth-fairness scheme soars rapidly, and reaches $5600$ seconds at $K^\textrm{SE}=15$.   
\section{Conclusion} \label{sec:conclusion}
In this work, we propose a joint framework to optimize the total communication time for a compression-aided FL algorithm. We have employed the coalition game framework as a simple and straight forward method to control the number of participating devices and bandwidth allocation. In future work, we will develop a reinforcement-learning solution to the considered problem, which is can adapt to any stochastic IoT networks. 

\bibliographystyle{IEEEtran}
\bibliography{Citation}

\end{document}